\documentclass{article}
\pdfoutput=1
\PassOptionsToPackage{numbers}{natbib}

    \usepackage[preprint]{neurips_2024}

\usepackage[utf8]{inputenc} 
\usepackage[T1]{fontenc}    
\usepackage{hyperref}       
\usepackage{url}            
\usepackage{booktabs}       
\usepackage{amsfonts}       
\usepackage{nicefrac}       
\usepackage{microtype}      
\usepackage{xcolor}         
\usepackage{graphicx}
\usepackage{algorithm}
\usepackage{algpseudocode}
\usepackage[normalem]{ulem}
\usepackage{wrapfig}
\usepackage{multicol}
\usepackage{listings}
\usepackage{subcaption}
\usepackage{makecell}
\usepackage{multirow}

\title{Effective Large Language Model Debugging with Best-first Tree Search}

\author{%
  Jialin Song \\
  NVIDIA \\
  \And
  Jonathan Raiman \\
  NVIDIA \\
  \AND
  Bryan Catanzaro \\
  NVIDIA \\
}

\begin{document}
\newcommand{\concat}{\mathbin\Vert}
\newcommand{\algorithmautorefname}{Algorithm}
\newcommand{\suggest}[2]{{\textbf{\sout{#1}}}\ {\color{blue} #2}}
\maketitle

\begin{abstract}
Large Language Models (LLMs) show promise in code generation tasks. However, their code-writing abilities are often limited in scope: while they can successfully implement simple functions, they struggle with more complex tasks. A fundamental difference with how an LLM writes code, compared to a human programmer, is that it cannot consistently spot and fix bugs. Debugging is a crucial skill for programmers and it enables iterative code refinement towards a correct implementation.
In this work, we propose a novel algorithm to enable LLMs to debug their code via self-reflection and search where a model attempts to identify its previous mistakes. Our key contributions are 1) a best-first tree search algorithm with self-reflections (BESTER) that achieves state-of-the-art Pass@1 in three code generation benchmarks. BESTER maintains its superiority when we measure pass rates taking into account additional inference costs incurred by tree search. 2) A novel interpretability study on what self-reflections attend to in buggy programs and how they impact bug fixes, which provides a deeper understanding of the debugging process. 3) An extensive study on when self-reflections are effective in finding bugs. 
\end{abstract}

\section{Introduction}
\label{sec:intro}
Large Language Models (LLMs) have demonstrated impressive capabilities in writing code from natural language task descriptions \citep{chen2021evaluating, achiam2023gpt, li2023starcoder, roziere2023code, guo2024deepseek}. Github Copilot has been shown to boost expert programmers' productivity \citep{dakhel2023github}. Code generation quality is a popular metric in evaluating foundation models \citep{touvron2023llama, team2023gemini, achiam2023gpt, parmar2024nemotron, jiang2024mixtral}.
However, most existing code generation evaluations ask a model to generate a single response to a question prompt, which represents a simplistic way to measure code writing ability as LLMs are presently unreliable at zero-shot code generation. We observe that code writing is an iterative process where human programmers alternate between writing code, testing code (with unit tests), and debugging test failures; whereas a language model would fail to generate proper code simply because a single wrongly predicted token due to stochastic sampling.

Inspired by the iterative nature of human programming, we ask: will an LLM finish more coding tasks if we allow it to iteratively debug its implementations using unit test feedback? In this paper, we answer in the affirmative by proposing a general best-first tree search framework that combines execution feedback and self-reflection and improves code generation abilities of LLMs. \textit{Self-reflection} refers to a special generation task where an LLM reviews its earlier outputs and attempts to identify logical flaws within. In code generation, we call self-reflection the process of prompting an LLM to produce debugging instructions and feedback when given as input the source code of a buggy program with the associated execution trace/errors/test results. Self-reflections can instruct LLMs to repair a buggy program, a process we call \textit{program repair}.
Our search algorithm, \underline{BE}st \underline{S}elf-reflection \underline{T}ree s\underline{E}a\underline{R}ch (BESTER), prompts a model to produce multiple self-reflections to identify issues in a buggy program and perform program repairs \citep{goues2019automated} based on each self-reflection, then it selects the repair that passes the highest number of passing (\textit{best repair}) to continue the search until finding a correct program or running out of search budget. We evaluate BESTER on language models of a range of sizes and capabilities across three popular code generation benchmarks, HumanEval \citep{chen2021evaluating}, MBPP \citep{austin2021program} and APPS \citep{hendrycks2021measuring},  and achieve state-of-the-art Pass@1 results. We also compare pass rate (\textit{Pass@Infer} defined in \autoref{sec:experiments:setup}) in the \textit{equal compute setting} (control for the same number of model inferences across each proposed approach), similar to the experiments in \cite{olausson2023self}. In the equal compute setting, BESTER also achieves superior Pass@Infer results. BESTER showcases an improved method to allocate inference-time compute to solve more coding tasks. 

Self-reflections play a central role in BESTER and we provide novel interpretability analyses focusing on two questions: 1) how buggy programs influence self-reflection generations and 2) the effect of different self-reflection properties on program repair success rate. To the best of our best knowledge, this is the first interpretability analysis for LLM code generation. We perform feature perturbations \citep{miglani2023using} to measure input importance with attribution scores. We focus the first part of our analysis on self-reflection generation and make the observation that self-reflections target specific lines in buggy programs. This line attribution specificity during self-reflection generation could be leveraged for more targeted repairs. In program repair generation, we find that the changed lines (the \textit{diff} in software development terms) have high scores attributed to self-reflection inputs, which implies that self-reflections cause the language models writing repairs to make targeted edits. 

In summary, our main contributions are:
\begin{itemize}
    \item BESTER, a best-first tree search algorithm that follows the most effective self-reflections and achieves state-of-the-art Pass@1 results in three programming benchmarks.
    \item A novel interpretability study on program repair and self-reflection generations to understand how LLMs approach debugging tasks.
    \item A detailed analysis on the impact of self-reflection quality on program repair success rate, and ablation studies on the effect of tree search parameter choices on downstream task performance.
\end{itemize}
\section{Related Works}
\label{sec:related}
\subsection{LLMs for Code Generation}
LLMs that are capable of generating code from natural language description are ubiquitous \citep{austin2021program, chen2021evaluating, li2022competition, roziere2023code, li2023starcoder, achiam2023gpt, team2023gemini, guo2024deepseek}. We often evaluate a code generation LLM by prompting it with a task description and evaluate the correctness of its output by executing a suite of input-output tests \citep{chen2021evaluating, austin2021program, hendrycks2021measuring}. Several recent works implemented search procedures with test execution information and model self-reflections to improve generation quality. Reflexion \citep{shinn2023reflexion} and self-refine \citep{madaan2023self} utilize model self-reflection to correct prior buggy implementations. LATS \citep{zhou2023language} uses Monte-Carlo tree search to search for a correct implementation. There is also recent work on improving self-reflection generation quality by finetuning models \citep{zheng2024opencodeinterpreter}. Our work proposes an alternative tree search algorithm based on best-first tree search with self-reflections.

Beyond small programming task benchmarks, researchers also developed larger scale programming tasks to capture real-world use cases. SWE-Bench \citep{jimenez2023swe} challenges LLMs to solve Github issues. DevBench \citep{li2024devbench} evaluates LLMs on different tasks in software development, such as project design and unit testing. In addition to code writing ability, retrieval-augmented generation (RAG) is often essential in accomplishing these tasks \citep{jimenez2023swe}.

\subsection{Logical Reasoning in LLMs}
Code generation belongs to the category of logical reasoning tasks, which includes mathematical reasoning \citep{qiao2022reasoning,toshniwal2024openmath, trinh2024solving}. Existing research works have developed many useful inference techniques to boost a model's reasoning ability, such as chain-of-thought prompting \citep{wei2022chain}, tree-of-thoughts \citep{yao2024tree}, and code interpreter integration \citep{imani2023mathprompter}. Least-to-most prompting \citep{zhou2022least} takes a divide-and-conquer approach to helping LLMs tackle complex problems.
Tree search is another popular algorithmic framework to iteratively build a correct reasoning path \citep{zhou2023language, trinh2024solving}. However, these techniques are fundamentally limited by a model's capacity to identify logical errors \citep{tyen2023llms}. Collecting more training data on correct reasoning paths to fine-tune base models is a fruitful approach \citep{hendrycks2020measuring,cobbe2021training, hendrycks2021math, toshniwal2024openmath}. A related direction leverages LLMs to generate synthetic data based on existing logical reasoning datasets \citep{huang2022large,luo2023wizardmath} and fine-tune base models with those data to self-improve the models. Combining LLMs with classical symbolic reasoning aims to benefit from the strong language ability of LLMs and the rigor made available by symbolic solvers \citep{pan2023logic, yang2023coupling}.
In this work, we evaluate LLMs' ability to identify bugs in programs and present a detailed analysis on the LLMs' ability to reason about their own mistakes in code generation.
\section{Methodology}
\label{sec:method}

\begin{figure}[t]
    \centering
    \includegraphics[width=0.8\textwidth]{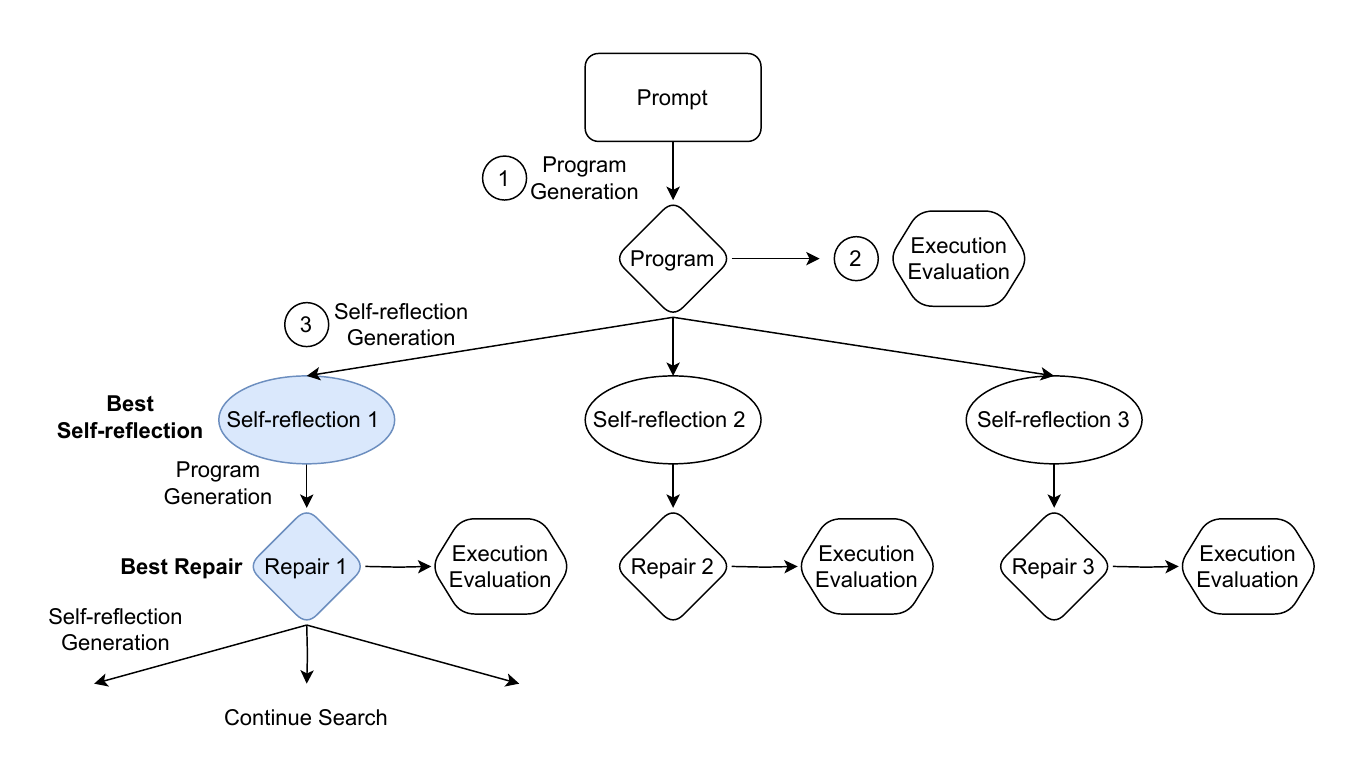}
    \caption{BESTER interleaves 1) program generation, 2) execution evaluation, and 3) self-reflection generation to refine programs. We sample multiple self-reflections based on a buggy program. Then an LLM generates a program repair based on each self-reflection. We select the repair with the highest score as the \textbf{Best Repair} and call its corresponding self-reflection the \textbf{Best Self-reflection} (highlighted in blue). If the best repair still fails some tests, we repeat self-reflection generations and program repairs until reaching a pre-determined maximal depth.}
    \label{fig:bester}
\end{figure}

In this section, we describe our proposed Best Self-reflection Tree Search (BESTER) algorithm for code generation. Each code generation problem $P$ contains a natural language specification $X$ on its functionality and a suite of test cases $T$ to evaluate the correctness of an implementation for $X$. After running a program $S$ on $T$, we record $r \in [0, 1]$, a score measuring the percentage of tests in $T$ that it passes. If $r = 1$, we find a program that passes all tests. Note that an important distinction exists between passing all tests and concluding that $S$ is correct according to the specification $X$: test coverage can be poor, allowing a program to be incorrect on corner cases outside the test suite but produce the desired outputs on the cases that were checked \citep{liu2023your}. To obtain a more truthful measure of correctness, we will evaluate with a more comprehensive test suite at the end of the tree search. We will provide more details in \autoref{sec:experiments}. 

Language model self-reflections are recently gaining popularity as an approach to refining its logical reasoning \citep{shinn2023reflexion, zhou2023language}. Broadly speaking, it is a technique to present a language model with an incorrect answer and prompt it to identify logical issues within. For code generation tasks, logical issues manifest as bugs that we wish for a model to rectify. Our approach incorporates self-reflections in a best-first tree algorithm to enable iterative solution refinement. BESTER interleaves three operations: program generation, execution evaluation, and self-reflection generation. 
\autoref{fig:bester} displays the core operations of our proposed algorithm. We describe each component in detail.

\paragraph{Program Generation} Given a problem $P = (X, T)$, we use a language model $LM$ to generate a program by greedy decoding $S = LM(X)$. While existing works show that temperature sampling improves code generation performances \citep{xu2022systematic,zhu2023improving}, we choose greedy decoding to better isolate the effect of searching with self-reflections. We prompt the model to follow a specific format of wrapping its code with \texttt{\textasciigrave\textasciigrave\textasciigrave} to simplify extraction. In the following tree search iterations, we prompt $LM$ with a self-reflection $SR$ and a buggy program $S$ to generate a new program repair $S' = LM(S \concat SR)$ via greedy decoding. We use $S \concat SR$ to represent a new prompt containing both $S$ and $SR$. 

\paragraph{Execution Evaluation} We measure the correctness of a program $S$ by running it with test cases in $T$ to compare output results, and obtain a score $r$. If $r = 1$, we find a program satisfying all test cases and terminate the search. If $r < 1$, we continue onto self-reflection generations.

\paragraph{Self-reflection Generation} If there are failed test cases, we emulate the debugging process by prompting a model to self-reflect on issues in $S$. We posit a successful debugger needs access to the problem specification $X$, the buggy program $S$, and the set of its failed test cases $T_{fail} \subset T$, thus we formulate a new prompt with these information $X \concat S \concat T_{fail}$. A model generates a self-reflection $SR = LM(X \concat S \concat T_{fail})$ to describe bugs in $S$ and potential actions to fix them. With $SR$ and the previous solution $S$, we ask the model to generate a program repair $S' = LM(S\concat SR)$.

\begin{algorithm}
\caption{Best Self-reflection Tree Search (BESTER)}
\label{alg:bester}
\begin{algorithmic}[1]
\State {\bf Input}: $LM$ a language model; $P=(X,T)$ a programming problem; $N$ max depth for tree search; $k$ the number of self-reflections to sample

\For{$i=1...N$}
\If {$i == 1$}
\State \texttt{prompts} = $[X]$ \Comment{First iteration uses the specification as prompt}
\Else 
\State \texttt{prompts = new\_prompts} \Comment{Later iterations use self-reflection for program repair}
\EndIf
\State $S = LM(\texttt{prompts})$ \Comment{Greedy decoding for each prompt}
\State Evaluate programs in $S$ with the test suite $T$
\State Select $S^*$ with the highest percentage of passing tests
\If {$S^*$ passes all tests}
\State return $S^*$
\EndIf
\State Format self-reflection prompt $X\concat S^*\concat T_{fail}$, where $T_{fail}$ is the set of failing tests by $S^*$
\State Sample $k$ self-reflections $SR = LM(X\concat S^*\concat T_{fail})$
\State \texttt{new\_prompts} $= []$
\For {$j=1...k$}
\State \texttt{new\_prompts}.append($S^*\concat SR[j]$) \Comment{Program repair prompt with self-reflection}
\EndFor
\EndFor
\State return $S^*$ \Comment{No program passing all tests is found, return the best we found}
\end{algorithmic}
\end{algorithm}

\paragraph{Best Self-reflection Tree Search (BESTER)} \autoref{alg:bester} describes our proposed best-first tree search with self-reflections. On lines 9 and 10, we evaluate programs with tests $T$ and select the one $S^*$ passing the most tests. Note for the first iteration, we only generate one program via greedy decoding and in the subsequent iterations, we greedily decode a program repair based on each self-reflection. On lines 14 and 15, we formulate the self-reflection prompt for $S^*$ and uses temperature sampling to generate $k$ different self-reflections. Temperature sampling helps the language model generate a diverse set of responses so it increases the chance of identifying bugs. After sampling self-reflections, we collect them into a set of new prompts with $S^*$ and each bug-identifying self-reflection (lines 16 - 19). The language model will attempt program repair in the beginning of the next iteration following the new prompts. We include the prompt templates in \autoref{sec:prompts}.

BESTER contains two important parameters: $N$ the max depth for tree search, controlling how deep the tree search grows, and $k$ the number of self-reflections to sample in each iteration, controlling the tree width. In \autoref{sec:experiments:ablation}, we provide an ablation study on the downstream performance impact of changing these two parameters.
\section{Experiments}
\label{sec:experiments}
We evaluate BESTER on three code generation benchmarks: HumanEval \citep{chen2021evaluating}, MBPP \citep{austin2021program} and APPS\citep{hendrycks2021measuring}. We investigate the following questions: 1) Is best-first search with self-reflection an effective algorithm at solving code generation tasks? (\autoref{sec:experiments:main}); 2) How do design decisions in BESTER influence its performances? (\autoref{sec:experiments:ablation}); 3) Can we interpret the generation process of self-reflections and program repairs? (\autoref{sec:experiments:interpret})
\subsection{Experiment Setup}
\label{sec:experiments:setup}
\paragraph{Models} We evaluate BESTER with three models: GPT-4\footnote{We use gpt-4-0613 in our experiments} \citep{achiam2023gpt}, deepseek-coder-33b-instruct (Deepseek) \citep{guo2024deepseek} and Meta-Llama-3-70B-Instruct (Llama3) \citep{llama3modelcard}. All three models are finetuned for instruction following, which is an important ability because in the self-reflection generation and program repair steps we need the model to carry out specific actions.

\paragraph{Datasets} We evaluate the correctness of a program with unit tests. For HumanEval and MBPP, we use the expanded test suites in EvalPlus \citep{evalplus}. 
For each problem, we divide its test suite into two subsets: a small subset for execution evaluation during search and a large one that acts as a held-out set to test the returned program from search. In our experiments, we label a program correct if it passes all cases in the held-out test set to ensure the separation between search and final evaluation. For HumanEval and MBPP, we use test cases in the original dataset during search and the rest of cases in EvalPlus as held-out sets. For APPS, we select 300 problems with at least 10 test cases. We use 3 tests during search and the rest as the held-out set.

\paragraph{Metrics} Pass@$k$ is the canonical metric to evaluate code generation tasks \citep{chen2021evaluating}. It measures the probability that at least one program from $k$ i.i.d. samples is correct. This metric is proposed for the sampling-based generation procedure but we can also apply it for search-based generation because of the creation of held-out test sets. We ensure that during search the model will not see test cases that the final evaluation is based on. When a search-based method, such as BESTER, only returns \textbf{a single} solution to evaluate on the held-out set, then Pass@1 is a valid metric. We remark that search-based generation such as \autoref{alg:bester} may require multiple model inferences to finish, but the Pass@1 metric ignores the additional computational cost. We address this limitation with a second metric \textit{Pass@Infer} that integrates the computation cost increase by measuring the test suite pass rate given a fixed number of model inference calls.

\paragraph{Baselines} We compare with three baseline methods. The first one is temperature sampling which does not utilize any code execution feedback. The second is best-first tree search with only execution feedback without self-reflections (Ex. Feedback). The last one is Reflexion \citep{shinn2023reflexion} which is a special case of BESTER if we only sample a single self-reflection at each iteration.

\paragraph{BESTER Parameters} We run BESTER with 2 rounds of self-reflections and sample 5 self-reflections each round with a 0.8 temperature term. This corresponds to setting $N=2, k=5$ in \autoref{alg:bester}. We use greedy decoding for both the initial and subsequent repair code generation. We provide an ablation on the choice of $N$ and $k$ in \autoref{sec:experiments:ablation}. With this parameter setting, we allocate 21 inference calls during tree search, 1 for the initial greedy decoding and $N \times k = 10$ each for self-reflection and program repair generation. So for Pass@Infer results in the next section, we fix number the inferences to be 21 for all methods.

\subsection{Main Results}
\label{sec:experiments:main}
\begin{table}[t]
\centering
\begin{tabular}{ccccc}
 \multirow{ 2}{*}{Model} & \multirow{ 2}{*}{Method} & \multicolumn{3}{c}{Pass@1 / Pass@Infer} \\ 
 & & HumanEval & MBPP & APPS \\ \hline
 GPT-4 & Sampling & 78.25\% / 82.67\% & 75.66\% / 78.30\% & 60.59\% / 67.85\% \\
 GPT-4 & Ex. Feedback & \textbf{90.85\%} / 88.41\% & 81.22\% / 80.69\% & 71.67\% / 69.33\% \\
 GPT-4 & Reflexion & 89.63\% / 89.02\% & 78.04\% / 78.04\% & 70.67\% / 69.33\% \\
 GPT-4 & BESTER & 90.24\% / \textbf{90.24\%} & \textbf{81.48\%} / \textbf{81.48\%} & \textbf{72.33\%} / \textbf{72.33\%} \\ \hline
 Deepseek & Sampling & 65.71\% / 74.46\% & 67.70\% / 73.41\% & 45.22\% / \textbf{57.54\%} \\
 Deepseek & Ex. Feedback & 76.22\% / 73.78\% & 75.93\% / 75.93\% & 53.67\% / 53.67\%\\
 Deepseek & Reflexion & 75.61\% / 73.17\% & 76.46\% / 76.46\% & 50.67\% / 50.67\% \\
 Deepseek & BESTER & \textbf{81.71\%} / \textbf{81.71\%} & \textbf{77.51\%} / \textbf{77.51\%} & \textbf{54.67\%} / 54.67\% \\ \hline
 Llama3 & Sampling & 71.17\% / 75.38\% & 68.25\% / 73.14\% & 48.40\% / 53.78\% \\
 Llama3 & Ex. Feedback & 79.27\% / 79.27\% & 72.49\% / 71.96\% & 58.00\% / 57.67\% \\
 Llama3 & Reflexion & 79.88\% / 79.88\% & 72.49\% / 72.49\% & 57.67\% / 57.67\% \\
 Llama3 & BESTER & \textbf{82.93\%} / \textbf{82.93\%} & \textbf{74.07\%} / \textbf{74.07\%} & \textbf{59.67\%} / \textbf{59.67\%} \\ \hline
 
\end{tabular}
\caption{Pass@1 and Pass@Infer results for three models across benchmarks. Higher is better. Bolded numbers represent best performing method for a given model. BESTER consistently outperforms baseline methods in both metrics.}
\label{experiments:table:main}
\end{table}

\begin{table}[t]
\centering
\begin{tabular}{ccccc}
 Model & Method & \multicolumn{3}{c}{Pass@1 / \# Problem} \\ 
 & & HumanEval & MBPP & APPS \\ \hline
 GPT-4 & Ex. Feedback & \textbf{70.59\%} / 17 & 34.29\% / 35 & 42.67\% / 75 \\
 GPT-4 & Reflexion & 63.16\% / 19 & 7.89\% / 38 & 38.46\% / 78 \\
 GPT-4 & BESTER & 69.57\% / 23 & \textbf{45.95\%} /37 & \textbf{45.21\%} / 73 \\ \hline
 Deepseek & Ex. Feedback & 25.00\% / 40 & 22.95\% / 61 & 18.85\% / 122 \\
 Deepseek & Reflexion & 22.50\% / 40 & 26.23\% / 61 & 11.48\% / 122 \\
 Deepseek & BESTER & \textbf{53.49\%} / 43 & \textbf{32.79\%} /61 & \textbf{21.31\%} / 122 \\ \hline
 Llama3 & Ex. Feedback & 33.33\% / 39 & 34.29\% / 70 & 22.12\% / 113 \\
 Llama3 & Reflexion & 35.90\% / 39 & 34.72\% / 72 & 21.24\% / 113 \\
 Llama3 & BESTER & \textbf{48.72\%} / 39 & \textbf{42.86\%} /70 & \textbf{26.44\%} / 113 \\ 
\end{tabular}
\caption{Search method Pass@1 results among problems not solved by greedy decoding. BESTER solves up to 2x the number of hard problems.}
\label{experiments:table:nongreedy}
\end{table}

\begin{minipage}[c]{0.47\textwidth}
\addtolength{\tabcolsep}{-0.3em}
\begin{tabular}{cccc}
 \multirow{ 2}{*}{Dataset} & \multicolumn{3}{c}{P(Correct $\vert$ Best)}\\
 & GPT-4 & Deepseek & Llama3 \\ \hline
HumanEval & 93.26\% & 55.73\% & 78.26\% \\ 
 MBPP & 84.29\% & 50.22\% & 66.06\% \\ 
 APPS & 98.51\% & 52.86\% & 87.71\%  \\ 
\end{tabular}
\captionof{table}{Relative frequency that a \textit{best self-reflection} is judged correct by GPT-4 from BESTER search. BESTER follows correct self-reflections in the majority of cases.}
\label{experiments:table:correct}
\end{minipage}
\hfill 
\noindent
\begin{minipage}[c]{0.49\textwidth}
\centering
\addtolength{\tabcolsep}{-0.3em}
\begin{tabular}{ccc}
 \multirow{ 2}{*}{Dataset} & \multicolumn{2}{c}{Pass rate} \\
 & Correct SR & Incorrect SR\\
 \hline
 HumanEval & 57.39\% & 20.00\% \\ 
 MBPP & 52.03\% & 25.00\% \\ 
 APPS & 45.74\% & 8.33\% \\ 
\end{tabular}
\captionof{table}{Correct self-reflections (SR), judged by GPT-4, are much more likely to lead to successful repairs than incorrect ones. These relative frequencies are estimated from BESTER with GPT-4 search trees.}
\label{experiments:table:success}
\end{minipage}

\autoref{experiments:table:main} shows Pass@1 and Pass@Infer results. BESTER is consistently the best performing method according to both metrics, achieving state-of-the-art results in all but one setting. In the equal compute setting, the strong performance of BESTER confirms that best-first tree search with self-reflections is an efficient method to consume model inferences to solve coding tasks. It is no surprise that the better the language model is, the higher the pass rates BESTER can achieve. Running BESTER with a weaker model, e.g., Deepseek, can obtain competitive Pass@1 results with GPT-4 using temperature sampling. Hence, BESTER holds promise to help close the gap between a weak and a strong model.

A subset of the problems in each dataset can be solved directly by the initial greedy decoding program from the problem prompt so we do not need to perform search for them. \autoref{experiments:table:nongreedy} shows Pass@1 results for problems where greedy decoding fails to solve them. This comparison highlights the search efficacy of BESTER as it solves up to 2x the number of these harder problems.

A key part of BESTER algorithm is the best-first selection rule to decide the next candidate program (best repair). We call a self-reflection \textit{best self-reflection} if it leads to a best repair. A natural question is how often best self-reflections correctly describe bugs. We design a custom prompt for GPT-4 to label a self-reflection correct if it is successful at describing bugs and incorrect otherwise. The relative frequency that best self-reflections are correct, P(Correct $\vert$ Best), is shown in \autoref{experiments:table:correct}. In all cases, the majority of best self-reflections are correct which implies BESTER is more likely to follow correct self-reflections that contain accurate debug information during tree search. Moreover, in head-to-head comparison between two models, the one with the higher P(Correct $\vert$ Best) achieves higher Pass@1 in 8 out of 9 cases. This is because following a correct self-reflection, the next program repair step is much more likely to yield a success, as shown in \autoref{experiments:table:success}, hence further improvement of BESTER requires advancement in more accurate self-reflection models.

\subsection{Ablation Study}
\label{sec:experiments:ablation}
In this section, we ablate two components in the BESTER algorithm: the selection rule during tree search (Line 10 in \autoref{alg:bester}), and inference allocations (values of $N, k$) in tree search.
\paragraph{Selection Rule} Selection rule determines which program repair to follow during tree search. We construct 4 heuristic selection rules: randomly select a repair (Random), select a repair generated from a self-reflection that contains code snippets (Code), select a repair having the minimal number of diffs from the buggy program (Min-diff), and select one having the maximal number of diffs (Max-diff).
\autoref{experiments:table:select} confirms that the number of passing tests is a more useful signal to guide the tree search than heuristic rules. We leave it as future work for better selection rules, for example taking the entire search tree into account in making selection decisions.

\begin{table}[t]
\centering
\begin{tabular}{*6c}
 Model  & Random & Code & Min-diff & Max-diff  & BESTER \\ \hline
 Deepseek & 75.61\% & 75.61\% & 70.73\% & 74.39\% & \textbf{81.71\%} \\ 
 Llama3 & 77.44\% & 79.27\% & 78.05\%  & 79.88\% & \textbf{82.32\%} \\ 
\end{tabular}
\caption{Pass@1 results for different selection rule in tree search on HumanEval. Best-first selection on the number of passing tests is crucial to achieving the highest Pass@1.}
\label{experiments:table:select}
\end{table}

\begin{figure}
\centering
\begin{subfigure}{.45\textwidth}
  \centering
  \includegraphics[width=\linewidth]{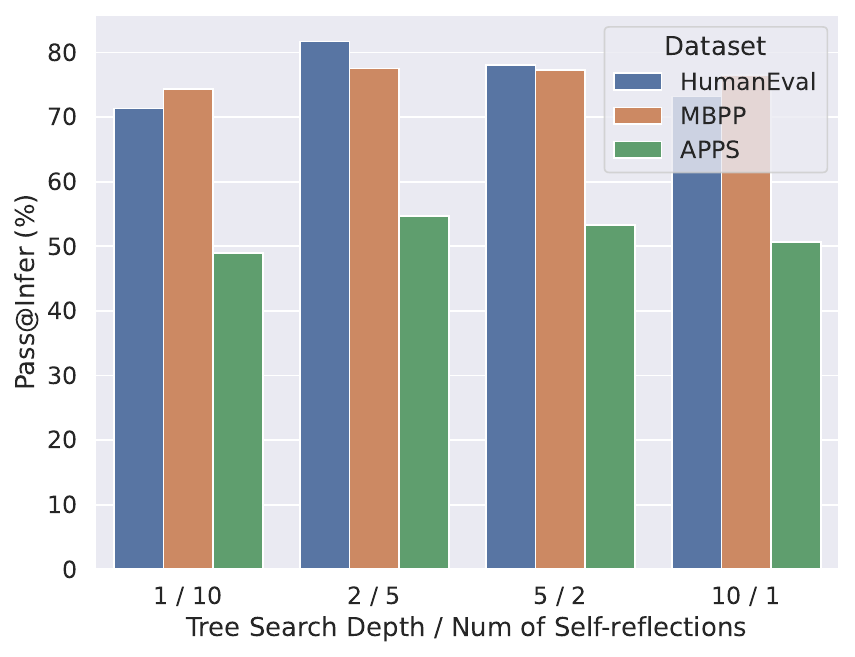}
  \caption{Deepseek}
  \label{fig:depth:deekseek}
\end{subfigure}%
\begin{subfigure}{.45\textwidth}
  \centering
  \includegraphics[width=\linewidth]{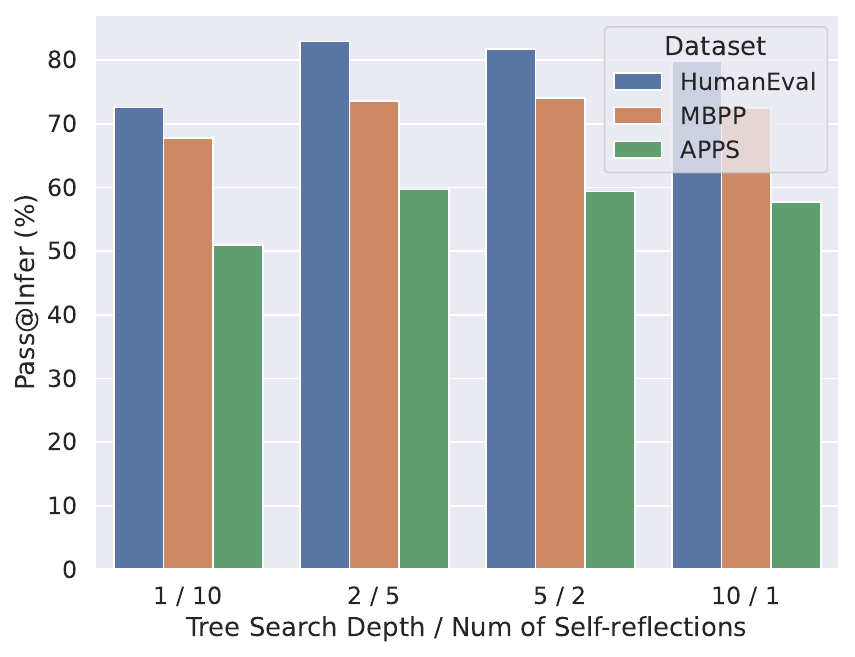}
  \caption{Llama3}
  \label{fig:depth:llama3}
\end{subfigure}
\caption{Pass@Infer results with different tree search depth and number of self-reflection configurations. Empirical results suggest choosing a depth of 2 and sampling 5 self-reflections at each step.}
\label{fig:depth}
\end{figure}

\paragraph{Inference Allocation} Two parameters in the BESTER algorithm control the shape of a search tree: $N$ the maximal depth to search and $k$ the number of self-reflections to sample at each iteration. In the equal compute setting, using a larger value of $N$ necessitates a smaller value of $k$ accordingly. Is it better to search with a shallow and wide tree or a deep and narrow one? We use the same 21 total inferences as in our main experiments. Subtracting the initial greedy decoding inference, we have 20 left for tree search. A program repair call follows each self-reflection generation so we can generate $20 / 2 = 10$ self-reflections during a tree search. Thus, we keep $N \times k = 10$ in this ablation study, and study 4 configurations $(N, k) = (1, 10), (2, 5), (5, 2), (10, 1)$.
\autoref{fig:depth} shows Pass@Infer comparisons. Configuration values in the middle tend to be better than the extremes. We find that $N = 2$ is empirically the best value. We leave adaptive setting values of $N$ and $k$ as future work.

\subsection{Model Interpretations}
\label{sec:experiments:interpret}
In this section, we present novel interpretability studies into two components in BESTER: self-reflection generation and program repair generation. We use Captum \citep{kokhlikyan2020captum} to perform input feature perturbation \citep{miglani2023using} to compute attribution scores. We use the log probability of a target string from the language model, given an input prompt, to compute attribution scores. 
Mathematically, the attribution score for an input-output pair ($I, O$) is $\log P_{LM(I)}(O) - \log P_{LM(I')}(O)$ where $P_{LM}(I)$ is the probability distribution over output sequences when a model $LM$ receives an input prompt $I$, and $I'$ is a perturbed prompt from $I$. It measures how much \textbf{more} likely a model generates $O$ when the input is $I$ compared to $I'$. A high attribution score means that the generation of $O$ depends heavily on $I$. Captum computes per output token attribution scores so that we can aggregate scores for parts of the output, for example, individual lines in a program.

We extract triples $(S, SR, S')$ from BESTER search trees, where $S$ is a buggy program, $SR$ is a self-reflection based on $S$, and $S'$ is a program repair based on $S$ and $SR$. We first use $(I, O) = (S, SR)$ to study how self-reflections depend on buggy programs. Then we use $(I, O) = (SR, S')$ to study how program repairs depend on self-reflections.

\begin{figure}
    \centering
    \begin{subfigure}[b]{\textwidth}
    \includegraphics[width=\textwidth]{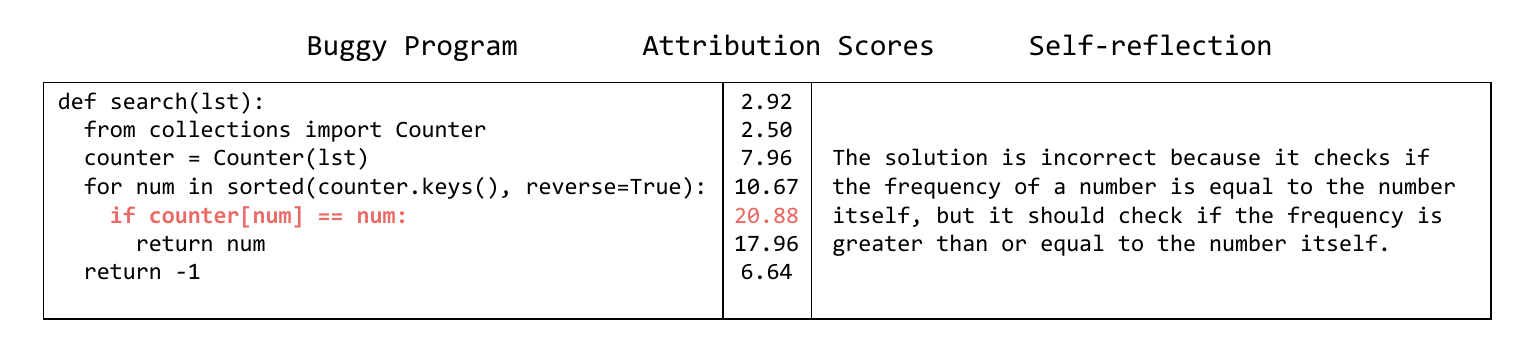}
    \caption{Attribution scores the self-reflection \textbf{attributes to} each line in the buggy program. The diff is between the buggy program and its repair (Repair Program in the example below). This self-reflection depends most heavily on this diff line.}
    \label{fig:reflection}
    \end{subfigure}

    \begin{subfigure}[b]{\textwidth}
    \includegraphics[width=\textwidth]{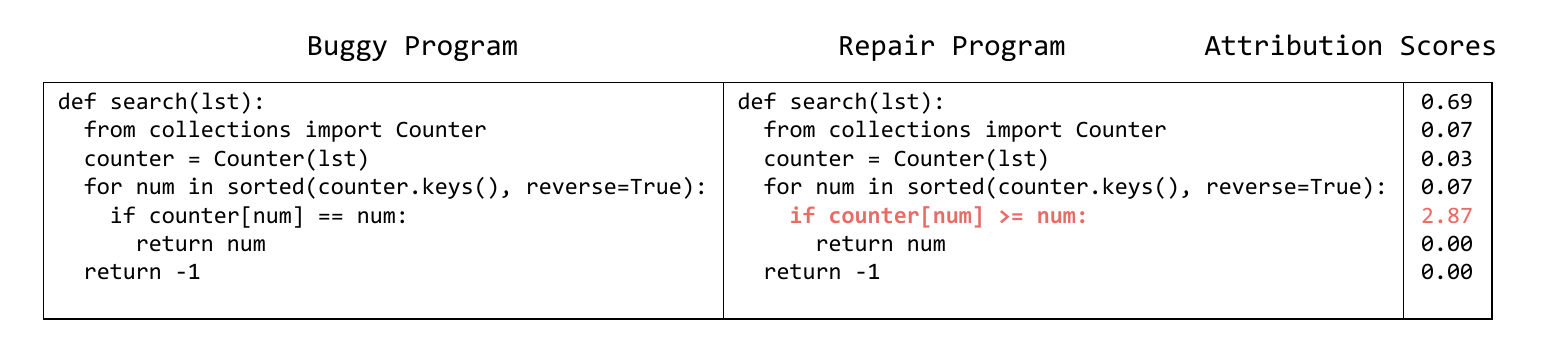}
    \caption{Attribution scores for each line in the repair program \textbf{attributed to} the self-reflection. A high score means the self-reflection heavily influences its generation. The diff line has the highest attribution score, so self-reflection causes this line change.}
    \label{fig:repair}
    \end{subfigure}
    \label{fig:attr_example}
    \caption{An example self-reflection and repair program from BESTER with Deepseek. We highlight the diff line in red.}
\end{figure}

\begin{figure}
\centering
\begin{subfigure}{.5\textwidth}
  \centering
  \includegraphics[width=\linewidth]{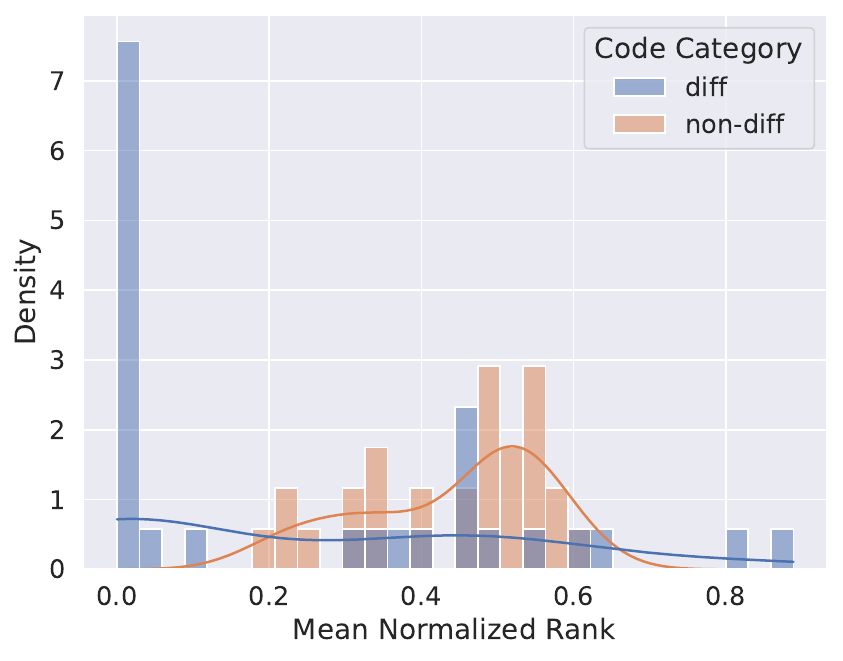}
  \caption{Deepseek on HumanEval}
\end{subfigure}%
\begin{subfigure}{.5\textwidth}
  \centering
  \includegraphics[width=\linewidth]{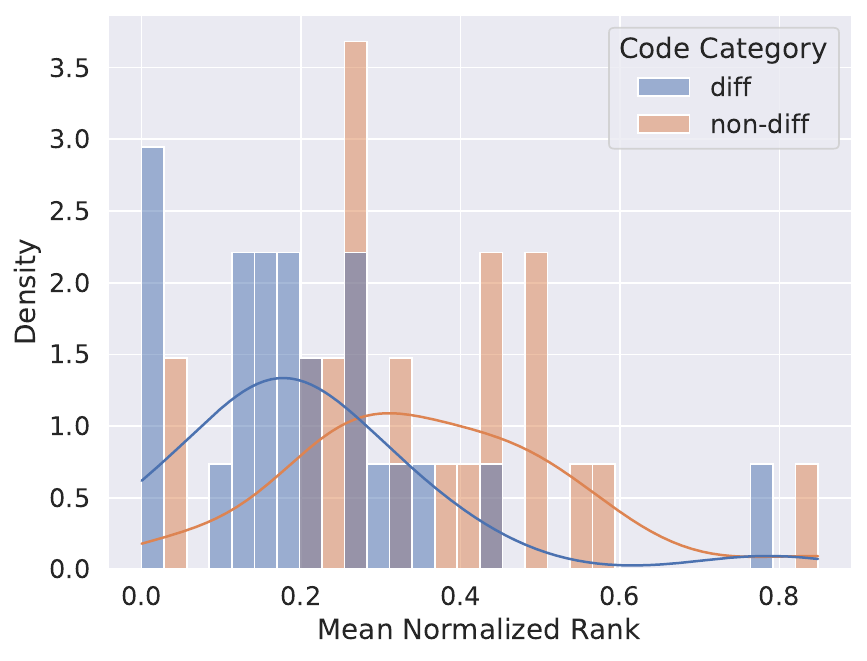}
  \caption{Llama3 on HumanEval}
\end{subfigure}
\caption{Mean normalized rank distributions for attribution scores that measure how much self-reflections depend on buggy programs. Lower values represent higher attribution scores. Lines that will be changed in the edited program have more influence on the self-reflection than lines that will remain the same.}
\label{fig:attr_feedback}
\end{figure}

\paragraph{Self-reflections target specific lines in buggy programs} We first study how self-reflection generation depends on buggy programs. \autoref{fig:reflection} shows an example from the Deepseek model. The self-reflection on the right has an attribution score (the middle column) for each line of the input buggy program on the left. 
Based on this self-reflection, Deepseek generates a repair program \autoref{fig:repair}.
We highlight in red the diff in the buggy program from the repair. In this example, the diff line has the highest attribution score which means this self-reflection depends most heavily on this line. Indeed, the self-reflection directly refers to the incorrect frequency comparison in the highlighted line. 

We hypothesize that attribution scores of diff lines tend to be higher than those of non-diff lines. As attribution score ranges vary from problem to problem, we transform them into normalized ranks: first sort the scores in descending order then assign a line a normalized rank of $r / (R-1)$ if its attribution score ranking is $r$, where $R$ is the total number of program lines. Then we compute the mean normalized ranks among diff and non-diff lines for each program, with results on HumanEval in
\autoref{fig:attr_feedback}. The mean normalized rank of diff lines is lower than that of non-diff for both Deepseek and Llama3, which implies that diff lines on average have higher attribution scores. The differences are statistically significance (\autoref{sec:interpret}). We carry out the same analyses for MBPP and APPS, and the same result holds. So we can conclude that self-reflections target diff lines in buggy programs.

\paragraph{Self-reflections cause targeted program edits} Next we study how self-reflections influence program repairs. We show an example in \autoref{fig:repair} which is a continuation from \autoref{fig:reflection}. For each line in the repair program, the rightmost column shows its corresponding attribution score on the self-reflection. 
Again, the diff line has the highest attribution score, which means that it depends most heavily on the self-reflection. Indeed, the self-reflection contains instruction on how to fix the bug.

\begin{figure}
\centering
\begin{subfigure}{.5\textwidth}
  \centering
  \includegraphics[width=\linewidth]{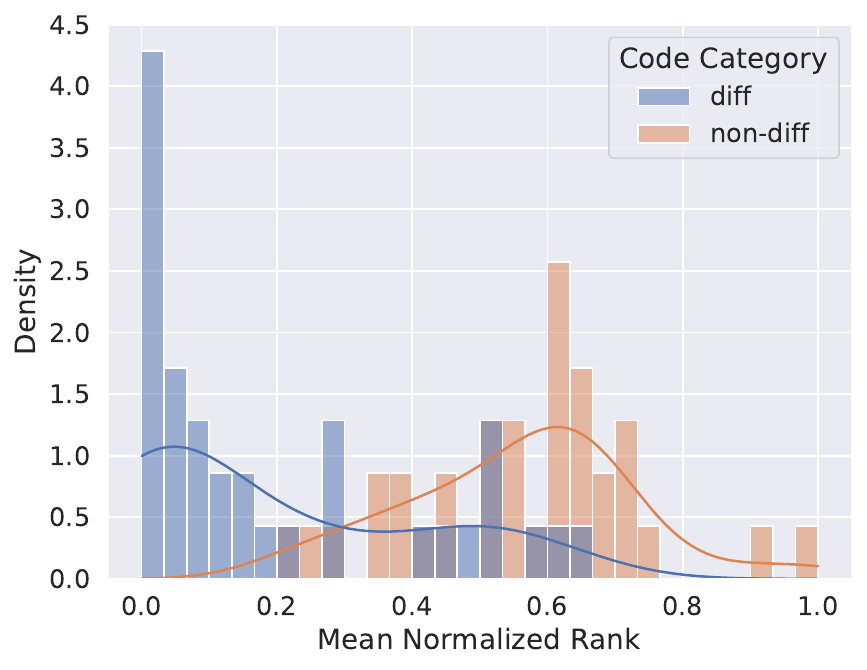}
  \caption{Deepseek on HumanEval}
\end{subfigure}%
\begin{subfigure}{.5\textwidth}
  \centering
  \includegraphics[width=\linewidth]{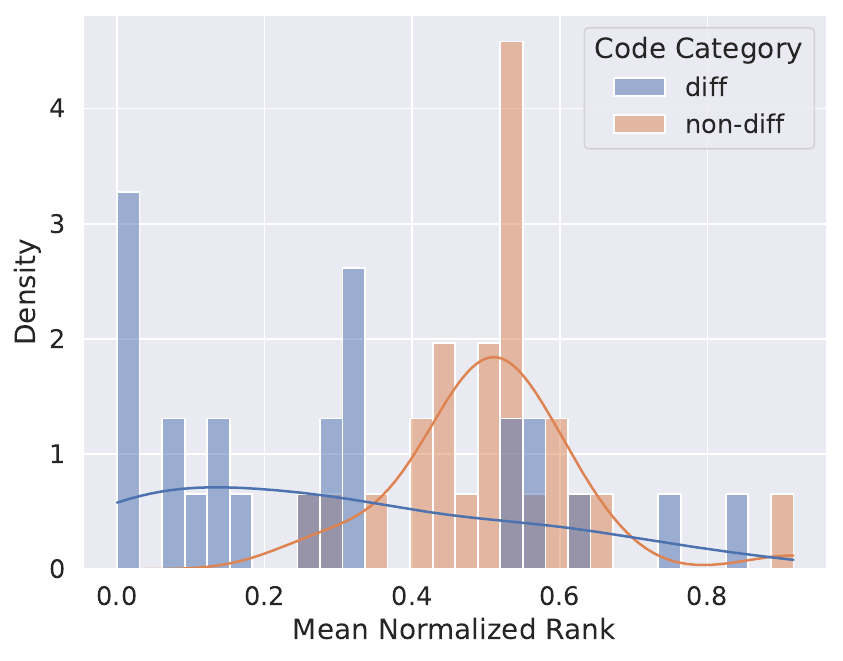}
  \caption{Llama3 on HumanEval}
\end{subfigure}
\caption{Mean normalized rank distributions attribution scores that measure how much repair programs depend on self-reflections. Lower values represent higher attribution scores. Self-reflections cause targeted code diff edits.}
\label{fig:attr_solution}
\end{figure}

\autoref{fig:attr_solution} compares mean normalized ranks of diff and non-diff lines on HumanEval. The mean normalized rank of diff lines is lower than that of non-diff lines for both Deepseek and Llama3, which implies that diff lines in general have higher attribution scores. So we can conclude that self-reflections cause targeted code diff edits.  

\section{Conclusion}
\label{sec:conclusion}
In this paper, we propose BESTER a best-first tree search algorithm with self-reflections for code generation tasks. This algorithm enables language models to solve programming tasks iteratively. BESTER achieves state-of-the-art Pass@1 using three different langauge models and excels in the equal compute setting as well. We analyze the connection between self-reflection correctness and its effectiveness in successful repairing a buggy program. Finally, we present a novel interpretability study on self-reflection and program repair generations.
Applying BESTER to other logical reasoning tasks is a natural next step. Real-world software engineering tasks pose a scalability challenge due to additional difficulties posed by the additional context required. Integrating BESTER into an agent framework is an interesting future direction. 

\bibliographystyle{unsrt}
\bibliography{main}

\begin{thebibliography}{10}

\bibitem{chen2021evaluating}
Mark Chen, Jerry Tworek, Heewoo Jun, Qiming Yuan, Henrique Ponde de~Oliveira Pinto, Jared Kaplan, Harri Edwards, Yuri Burda, Nicholas Joseph, Greg Brockman, et~al.
\newblock Evaluating large language models trained on code.
\newblock {\em arXiv preprint arXiv:2107.03374}, 2021.

\bibitem{achiam2023gpt}
Josh Achiam, Steven Adler, Sandhini Agarwal, Lama Ahmad, Ilge Akkaya, Florencia~Leoni Aleman, Diogo Almeida, Janko Altenschmidt, Sam Altman, Shyamal Anadkat, et~al.
\newblock Gpt-4 technical report.
\newblock {\em arXiv preprint arXiv:2303.08774}, 2023.

\bibitem{li2023starcoder}
Raymond Li, Loubna~Ben Allal, Yangtian Zi, Niklas Muennighoff, Denis Kocetkov, Chenghao Mou, Marc Marone, Christopher Akiki, Jia Li, Jenny Chim, et~al.
\newblock Starcoder: may the source be with you!
\newblock {\em arXiv preprint arXiv:2305.06161}, 2023.

\bibitem{roziere2023code}
Baptiste Roziere, Jonas Gehring, Fabian Gloeckle, Sten Sootla, Itai Gat, Xiaoqing~Ellen Tan, Yossi Adi, Jingyu Liu, Tal Remez, J{\'e}r{\'e}my Rapin, et~al.
\newblock Code llama: Open foundation models for code.
\newblock {\em arXiv preprint arXiv:2308.12950}, 2023.

\bibitem{guo2024deepseek}
Daya Guo, Qihao Zhu, Dejian Yang, Zhenda Xie, Kai Dong, Wentao Zhang, Guanting Chen, Xiao Bi, Y~Wu, YK~Li, et~al.
\newblock Deepseek-coder: When the large language model meets programming--the rise of code intelligence.
\newblock {\em arXiv preprint arXiv:2401.14196}, 2024.

\bibitem{dakhel2023github}
Arghavan~Moradi Dakhel, Vahid Majdinasab, Amin Nikanjam, Foutse Khomh, Michel~C Desmarais, and Zhen Ming~Jack Jiang.
\newblock Github copilot ai pair programmer: Asset or liability?
\newblock {\em Journal of Systems and Software}, 203:111734, 2023.

\bibitem{touvron2023llama}
Hugo Touvron, Louis Martin, Kevin Stone, Peter Albert, Amjad Almahairi, Yasmine Babaei, Nikolay Bashlykov, Soumya Batra, Prajjwal Bhargava, Shruti Bhosale, et~al.
\newblock Llama 2: Open foundation and fine-tuned chat models.
\newblock {\em arXiv preprint arXiv:2307.09288}, 2023.

\bibitem{team2023gemini}
Gemini Team, Rohan Anil, Sebastian Borgeaud, Yonghui Wu, Jean-Baptiste Alayrac, Jiahui Yu, Radu Soricut, Johan Schalkwyk, Andrew~M Dai, Anja Hauth, et~al.
\newblock Gemini: a family of highly capable multimodal models.
\newblock {\em arXiv preprint arXiv:2312.11805}, 2023.

\bibitem{parmar2024nemotron}
Jupinder Parmar, Shrimai Prabhumoye, Joseph Jennings, Mostofa Patwary, Sandeep Subramanian, Dan Su, Chen Zhu, Deepak Narayanan, Aastha Jhunjhunwala, Ayush Dattagupta, et~al.
\newblock Nemotron-4 15b technical report.
\newblock {\em arXiv preprint arXiv:2402.16819}, 2024.

\bibitem{jiang2024mixtral}
Albert~Q Jiang, Alexandre Sablayrolles, Antoine Roux, Arthur Mensch, Blanche Savary, Chris Bamford, Devendra~Singh Chaplot, Diego de~las Casas, Emma~Bou Hanna, Florian Bressand, et~al.
\newblock Mixtral of experts.
\newblock {\em arXiv preprint arXiv:2401.04088}, 2024.

\bibitem{goues2019automated}
Claire~Le Goues, Michael Pradel, and Abhik Roychoudhury.
\newblock Automated program repair.
\newblock {\em Communications of the ACM}, 62(12):56--65, 2019.

\bibitem{austin2021program}
Jacob Austin, Augustus Odena, Maxwell Nye, Maarten Bosma, Henryk Michalewski, David Dohan, Ellen Jiang, Carrie Cai, Michael Terry, Quoc Le, et~al.
\newblock Program synthesis with large language models.
\newblock {\em arXiv preprint arXiv:2108.07732}, 2021.

\bibitem{hendrycks2021measuring}
Dan Hendrycks, Steven Basart, Saurav Kadavath, Mantas Mazeika, Akul Arora, Ethan Guo, Collin Burns, Samir Puranik, Horace He, Dawn Song, et~al.
\newblock Measuring coding challenge competence with apps.
\newblock {\em arXiv preprint arXiv:2105.09938}, 2021.

\bibitem{olausson2023self}
Theo~X Olausson, Jeevana~Priya Inala, Chenglong Wang, Jianfeng Gao, and Armando Solar-Lezama.
\newblock Is self-repair a silver bullet for code generation?
\newblock In {\em The Twelfth International Conference on Learning Representations}, 2023.

\bibitem{miglani2023using}
Vivek Miglani, Aobo Yang, Aram~H Markosyan, Diego Garcia-Olano, and Narine Kokhlikyan.
\newblock Using captum to explain generative language models.
\newblock {\em arXiv preprint arXiv:2312.05491}, 2023.

\bibitem{li2022competition}
Yujia Li, David Choi, Junyoung Chung, Nate Kushman, Julian Schrittwieser, R{\'e}mi Leblond, Tom Eccles, James Keeling, Felix Gimeno, Agustin Dal~Lago, et~al.
\newblock Competition-level code generation with alphacode.
\newblock {\em Science}, 378(6624):1092--1097, 2022.

\bibitem{shinn2023reflexion}
Noah Shinn, Federico Cassano, Ashwin Gopinath, Karthik~R Narasimhan, and Shunyu Yao.
\newblock Reflexion: language agents with verbal reinforcement learning.
\newblock In {\em Thirty-seventh Conference on Neural Information Processing Systems}, 2023.

\bibitem{madaan2023self}
Aman Madaan, Niket Tandon, Prakhar Gupta, Skyler Hallinan, Luyu Gao, Sarah Wiegreffe, Uri Alon, Nouha Dziri, Shrimai Prabhumoye, Yiming Yang, et~al.
\newblock Self-refine: Iterative refinement with self-feedback.
\newblock In {\em Thirty-seventh Conference on Neural Information Processing Systems}, 2023.

\bibitem{zhou2023language}
Andy Zhou, Kai Yan, Michal Shlapentokh-Rothman, Haohan Wang, and Yu-Xiong Wang.
\newblock Language agent tree search unifies reasoning acting and planning in language models.
\newblock {\em arXiv preprint arXiv:2310.04406}, 2023.

\bibitem{zheng2024opencodeinterpreter}
Tianyu Zheng, Ge~Zhang, Tianhao Shen, Xueling Liu, Bill~Yuchen Lin, Jie Fu, Wenhu Chen, and Xiang Yue.
\newblock Opencodeinterpreter: Integrating code generation with execution and refinement.
\newblock {\em arXiv preprint arXiv:2402.14658}, 2024.

\bibitem{jimenez2023swe}
Carlos~E Jimenez, John Yang, Alexander Wettig, Shunyu Yao, Kexin Pei, Ofir Press, and Karthik Narasimhan.
\newblock Swe-bench: Can language models resolve real-world github issues?
\newblock {\em arXiv preprint arXiv:2310.06770}, 2023.

\bibitem{li2024devbench}
Bowen Li, Wenhan Wu, Ziwei Tang, Lin Shi, John Yang, Jinyang Li, Shunyu Yao, Chen Qian, Binyuan Hui, Qicheng Zhang, et~al.
\newblock Devbench: A comprehensive benchmark for software development.
\newblock {\em arXiv preprint arXiv:2403.08604}, 2024.

\bibitem{qiao2022reasoning}
Shuofei Qiao, Yixin Ou, Ningyu Zhang, Xiang Chen, Yunzhi Yao, Shumin Deng, Chuanqi Tan, Fei Huang, and Huajun Chen.
\newblock Reasoning with language model prompting: A survey.
\newblock {\em arXiv preprint arXiv:2212.09597}, 2022.

\bibitem{toshniwal2024openmath}
Shubham Toshniwal, Ivan Moshkov, Sean Narenthiran, Daria Gitman, Fei Jia, and Igor Gitman.
\newblock Openmathinstruct-1: A 1.8 million math instruction tuning dataset.
\newblock {\em arXiv preprint arXiv: Arxiv-2402.10176}, 2024.

\bibitem{trinh2024solving}
Trieu~H Trinh, Yuhuai Wu, Quoc~V Le, He~He, and Thang Luong.
\newblock Solving olympiad geometry without human demonstrations.
\newblock {\em Nature}, 625(7995):476--482, 2024.

\bibitem{wei2022chain}
Jason Wei, Xuezhi Wang, Dale Schuurmans, Maarten Bosma, Fei Xia, Ed~Chi, Quoc~V Le, Denny Zhou, et~al.
\newblock Chain-of-thought prompting elicits reasoning in large language models.
\newblock {\em Advances in neural information processing systems}, 35:24824--24837, 2022.

\bibitem{yao2024tree}
Shunyu Yao, Dian Yu, Jeffrey Zhao, Izhak Shafran, Tom Griffiths, Yuan Cao, and Karthik Narasimhan.
\newblock Tree of thoughts: Deliberate problem solving with large language models.
\newblock {\em Advances in Neural Information Processing Systems}, 36, 2024.

\bibitem{imani2023mathprompter}
Shima Imani, Liang Du, and Harsh Shrivastava.
\newblock Mathprompter: Mathematical reasoning using large language models.
\newblock {\em arXiv preprint arXiv:2303.05398}, 2023.

\bibitem{zhou2022least}
Denny Zhou, Nathanael Sch{\"a}rli, Le~Hou, Jason Wei, Nathan Scales, Xuezhi Wang, Dale Schuurmans, Claire Cui, Olivier Bousquet, Quoc Le, et~al.
\newblock Least-to-most prompting enables complex reasoning in large language models.
\newblock {\em arXiv preprint arXiv:2205.10625}, 2022.

\bibitem{tyen2023llms}
Gladys Tyen, Hassan Mansoor, Peter Chen, Tony Mak, and Victor C{\u{a}}rbune.
\newblock Llms cannot find reasoning errors, but can correct them!
\newblock {\em arXiv preprint arXiv:2311.08516}, 2023.

\bibitem{hendrycks2020measuring}
Dan Hendrycks, Collin Burns, Steven Basart, Andy Zou, Mantas Mazeika, Dawn Song, and Jacob Steinhardt.
\newblock Measuring massive multitask language understanding.
\newblock {\em arXiv preprint arXiv:2009.03300}, 2020.

\bibitem{cobbe2021training}
Karl Cobbe, Vineet Kosaraju, Mohammad Bavarian, Mark Chen, Heewoo Jun, Lukasz Kaiser, Matthias Plappert, Jerry Tworek, Jacob Hilton, Reiichiro Nakano, et~al.
\newblock Training verifiers to solve math word problems.
\newblock {\em arXiv preprint arXiv:2110.14168}, 2021.

\bibitem{hendrycks2021math}
Dan Hendrycks, Collin Burns, Saurav Kadavath, Akul Arora, Steven Basart, Eric Tang, Dawn Song, and Jacob Steinhardt.
\newblock Measuring mathematical problem solving with the math dataset.
\newblock {\em arXiv preprint arXiv:2103.03874}, 2021.

\bibitem{huang2022large}
Jiaxin Huang, Shixiang~Shane Gu, Le~Hou, Yuexin Wu, Xuezhi Wang, Hongkun Yu, and Jiawei Han.
\newblock Large language models can self-improve.
\newblock {\em arXiv preprint arXiv:2210.11610}, 2022.

\bibitem{luo2023wizardmath}
Haipeng Luo, Qingfeng Sun, Can Xu, Pu~Zhao, Jianguang Lou, Chongyang Tao, Xiubo Geng, Qingwei Lin, Shifeng Chen, and Dongmei Zhang.
\newblock Wizardmath: Empowering mathematical reasoning for large language models via reinforced evol-instruct.
\newblock {\em arXiv preprint arXiv:2308.09583}, 2023.

\bibitem{pan2023logic}
Liangming Pan, Alon Albalak, Xinyi Wang, and William~Yang Wang.
\newblock Logic-lm: Empowering large language models with symbolic solvers for faithful logical reasoning.
\newblock {\em arXiv preprint arXiv:2305.12295}, 2023.

\bibitem{yang2023coupling}
Zhun Yang, Adam Ishay, and Joohyung Lee.
\newblock Coupling large language models with logic programming for robust and general reasoning from text.
\newblock {\em arXiv preprint arXiv:2307.07696}, 2023.

\bibitem{liu2023your}
Jiawei Liu, Chunqiu~Steven Xia, Yuyao Wang, and LINGMING ZHANG.
\newblock Is your code generated by chatgpt really correct? rigorous evaluation of large language models for code generation.
\newblock In {\em Thirty-seventh Conference on Neural Information Processing Systems}, 2023.

\bibitem{xu2022systematic}
Frank~F Xu, Uri Alon, Graham Neubig, and Vincent~Josua Hellendoorn.
\newblock A systematic evaluation of large language models of code.
\newblock In {\em Proceedings of the 6th ACM SIGPLAN International Symposium on Machine Programming}, pages 1--10, 2022.

\bibitem{zhu2023improving}
Yuqi Zhu, Jia~Allen Li, Ge~Li, YunFei Zhao, Jia Li, Zhi Jin, and Hong Mei.
\newblock Improving code generation by dynamic temperature sampling.
\newblock {\em arXiv preprint arXiv:2309.02772}, 2023.

\bibitem{llama3modelcard}
AI@Meta.
\newblock Llama 3 model card.
\newblock 2024.

\bibitem{evalplus}
Jiawei Liu, Chunqiu~Steven Xia, Yuyao Wang, and Lingming Zhang.
\newblock Is your code generated by chat{GPT} really correct? rigorous evaluation of large language models for code generation.
\newblock In {\em Thirty-seventh Conference on Neural Information Processing Systems}, 2023.

\bibitem{kokhlikyan2020captum}
Narine Kokhlikyan, Vivek Miglani, Miguel Martin, Edward Wang, Bilal Alsallakh, Jonathan Reynolds, Alexander Melnikov, Natalia Kliushkina, Carlos Araya, Siqi Yan, and Orion Reblitz-Richardson.
\newblock Captum: A unified and generic model interpretability library for pytorch, 2020.

\end{thebibliography}
\newpage
\appendix
\section{Discussion}
\label{sec:impacts}
\paragraph{Limitations}
The main limitation in this study is the scale of studied code generation tasks as most of the solution programs have fewer than 100 lines of code. Real-life software engineering projects can easily reach tens of thousands of lines so whether BESTER can handle code generation tasks at a much larger scale is an open question. Another limitation is that BESTER is an inference-time search technique to enhance an existing model. One could argue that model finetuning is necessary to improve performance even further. We leave this as future direction.

\paragraph{Broader Impacts}
Automatic code generation with LLMs holds promise to boost the productivity for software engineers. Tools like Github Copilot are already being integrated into daily workflows. The usefulness of such tools hinge on whether they can consistently satisfy their users' needs. BESTER improves generation quality and could make those tools more helpful. However, there is always the possibility of misuse by malicious agents, e.g., to generate malware at scale. Even in legitimate use cases, automatic code generation tools can introduce bugs that are costly to fix. So we need to take a balanced view in order to improve code generation models. 

\section{Dataset Details}
\label{sec:dataset-detail}
In \autoref{sec:experiments}, we use three datasets: HumanEval \citep{chen2021evaluating}, MBPP \citep{austin2021program} and APPS \citep{hendrycks2021measuring}. We use test suites in EvalPlus \citep{evalplus} for HumanEval and MBPP final evaluations. EvalPlus suite covers all 164 problems in HumanEval and 378 problems in MBPP. We subsample 300 problems from the training split of APPS. We make sure each problem has at least 10 test cases and its type is call-based. 

The problem indices we use for APPS are: 3329, 4460, 2772, 3433, 4135, 3697, 1613, 3837, 4133, 3758, 3995, 4209, 3817, 3763, 3622, 4262, 3522, 4123, 1625, 4641, 1629, 4588, 2995, 3419, 4636, 2927, 2728, 4091, 4042, 4276, 3814, 2730, 3478, 3155, 4281, 2794, 4424, 4050, 3562, 2758, 4409, 4339, 2710, 2714, 3301, 2799, 4179, 3861, 3048, 3203, 2661, 4650, 4242, 3470, 3822, 4458, 2731, 4378, 2842, 4035, 3769, 2700, 4523, 3565, 4069, 3826, 4531, 3066, 3606, 3222, 2899, 2695, 4440, 3454, 3174, 2936, 4648, 4232, 2752, 3505, 3891, 3492, 4346, 4709, 4258, 3276, 3064, 2785, 4147, 4271, 4176, 3842, 4464, 4105, 3445, 2801, 3989, 4616, 2744, 4752, 3060, 4668, 3308, 2879, 4104, 2892, 4018, 3096, 3024, 2652, 3217, 4140, 4577, 4298, 2885, 4096, 4084, 2998, 3907, 4169, 4329, 2860, 3461, 3229, 2975, 3016, 4556, 4260, 2828, 4007, 2949, 4596, 2957, 4046, 4509, 3164, 3768, 3740, 3099, 2654, 2706, 3486, 2774, 4312, 4168, 4487, 4186, 3811, 2783, 3411, 4174, 4029, 3483, 4332, 3130, 3213, 3030, 2711, 3927, 2698, 3777, 4514, 3463, 3600, 3909, 3460, 4461, 1660, 3134, 3242, 4374, 3937, 3128, 3786, 3504, 4322, 3446, 4185, 4557, 3334, 4292, 3238, 3243, 3360, 1633, 4244, 3475, 1643, 1611, 4450, 4445, 4144, 3166, 2845, 2793, 3878, 4022, 4365, 3356, 3274, 3605, 4499, 4327, 3506, 3523, 3139, 3372, 3601, 2874, 3519, 3674, 3205, 4548, 4240, 2830, 3888, 3974, 3042, 3176, 4127, 3823, 3561, 2736, 1618, 1658, 3391, 3654, 3045, 4220, 3211, 3514, 4353, 4740, 4067, 1652, 4626, 4369, 3744, 3727, 4200, 3062, 3895, 3040, 2868, 4211, 4521, 2779, 4212, 4634, 4667, 3163, 4664, 4044, 4682, 4685, 3966, 4724, 3306, 4643, 4484, 3862, 3594, 3036, 3881, 3094, 3224, 4608, 4164, 4057, 4020, 2651, 4572, 4387, 3187, 4126, 4076, 4654, 2881, 1647, 3358, 3880, 3320, 4705, 4693, 1638, 2814, 3491, 4610, 3402, 3554, 4465, 2802, 3569, 3531, 3244, 3930, 3083, 4227, 4361, 3346.

\begin{figure}
\centering
\begin{subfigure}[b]{\textwidth}
\begin{subfigure}{.5\textwidth}
  \centering
  \includegraphics[width=\linewidth]{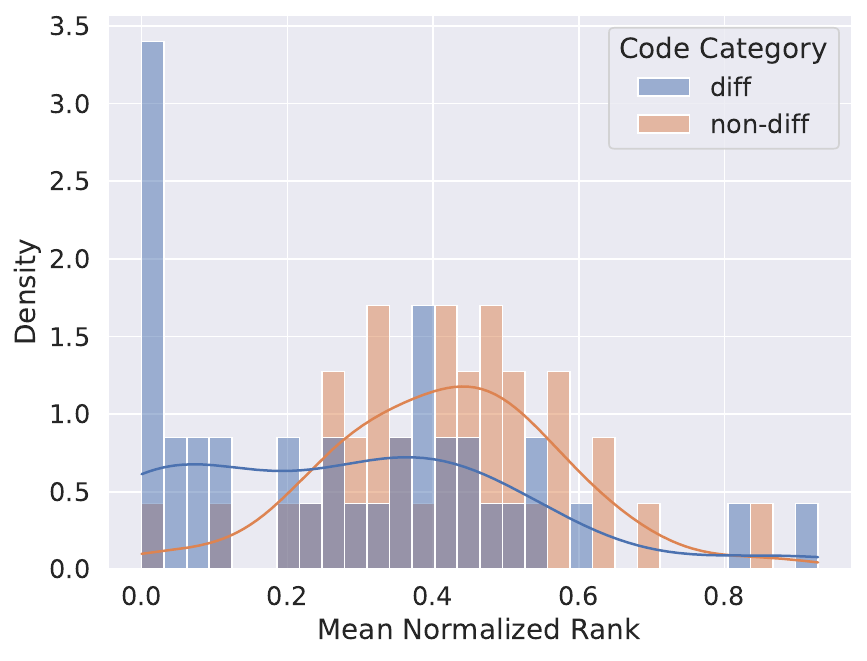}
  \caption{Deepseek on MBPP}
\end{subfigure}%
\begin{subfigure}{.5\textwidth}
  \centering
  \includegraphics[width=\linewidth]{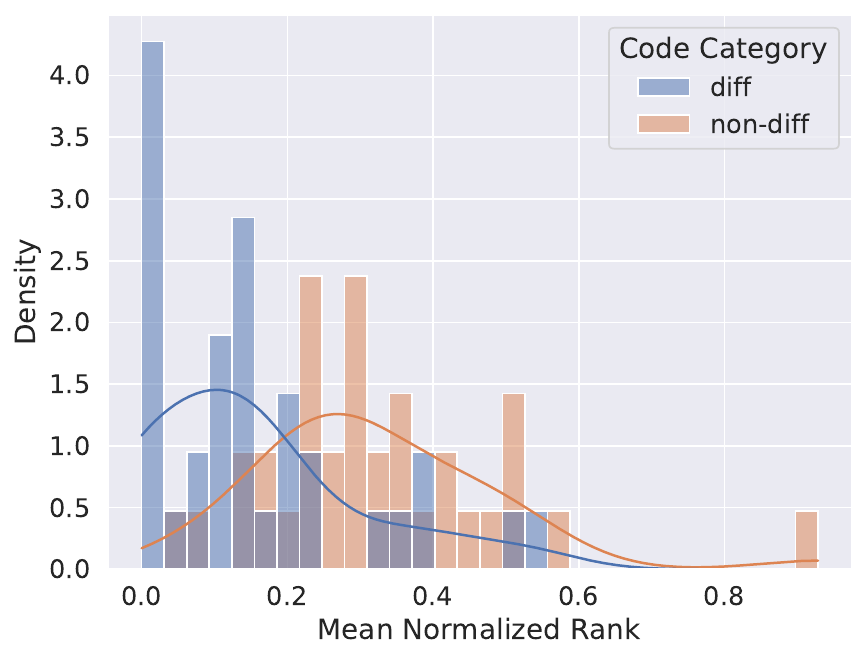}
  \caption{Llama3 on MBPP}
\end{subfigure}
\end{subfigure}

\begin{subfigure}[b]{\textwidth}
    
\begin{subfigure}{.5\textwidth}
  \centering
  \includegraphics[width=\linewidth]{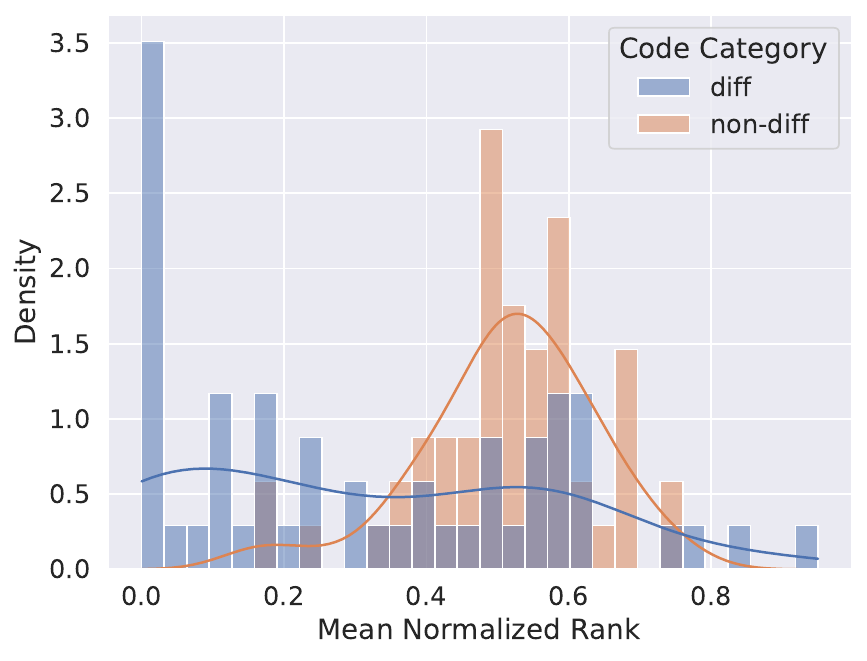}
  \caption{Deepseek on APPS}
\end{subfigure}%
\begin{subfigure}{.5\textwidth}
  \centering
  \includegraphics[width=\linewidth]{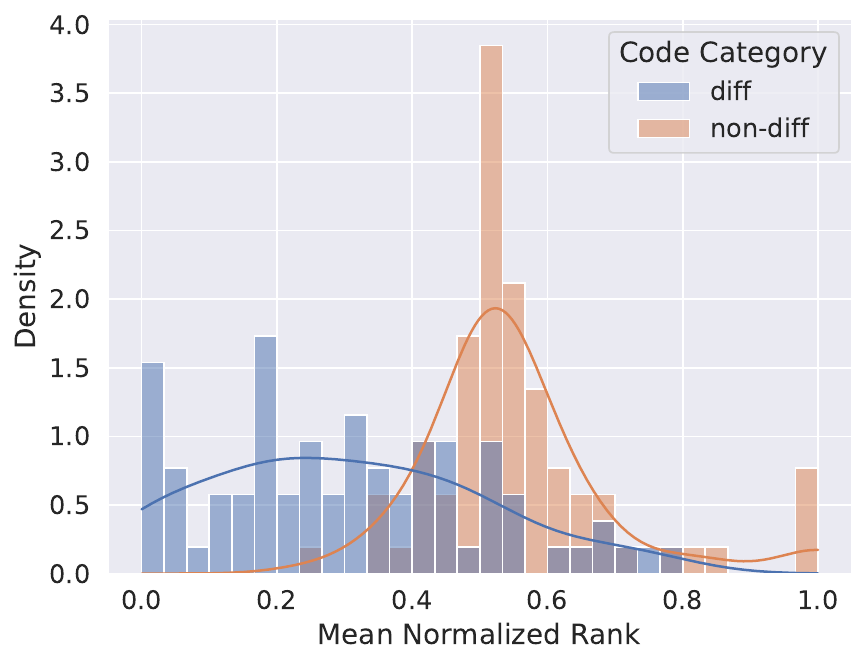}
  \caption{Llama3 on APPS}
\end{subfigure}
\end{subfigure}
\caption{Mean normalized rank distributions for attribution scores that measure how much self-reflections depend on buggy programs. Lower values represent higher attribution scores. Lines that will be changed in the edited program have more influence on the self-reflection than lines that will remain the same.}
\label{fig:attr_feedback:other}
\end{figure}

\begin{figure}
\centering
\begin{subfigure}[b]{\textwidth}
\begin{subfigure}{.5\textwidth}
  \centering
  \includegraphics[width=\linewidth]{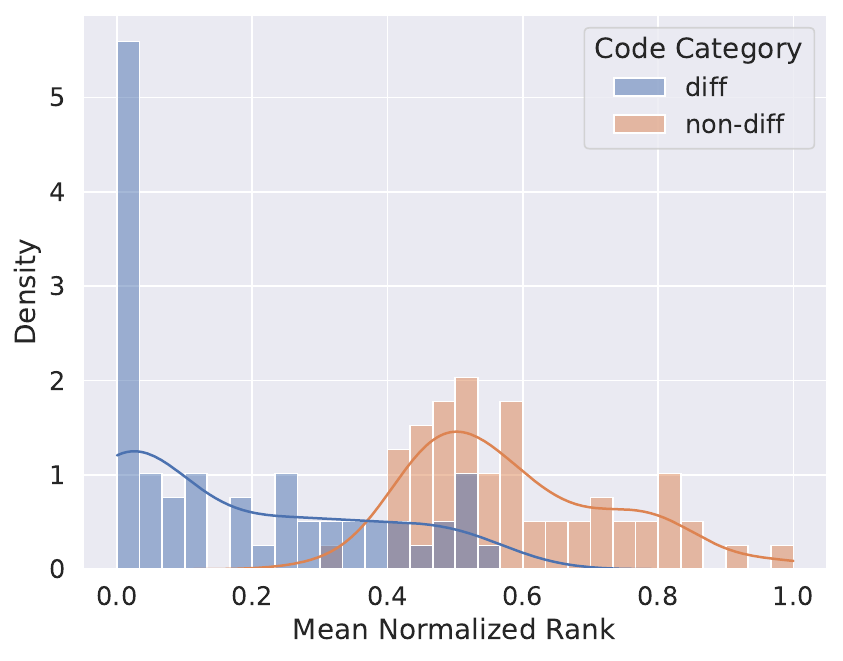}
  \caption{Deepseek on MBPP}
\end{subfigure}%
\begin{subfigure}{.5\textwidth}
  \centering
  \includegraphics[width=\linewidth]{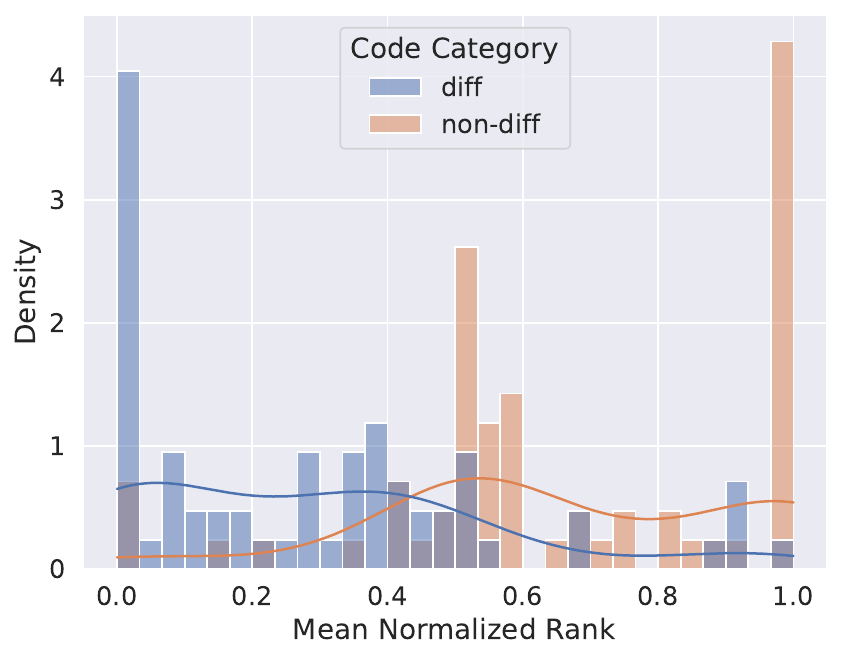}
  \caption{Llama3 on MBPP}
\end{subfigure}
\end{subfigure}

\begin{subfigure}[b]{\textwidth}
    
\begin{subfigure}{.5\textwidth}
  \centering
  \includegraphics[width=\linewidth]{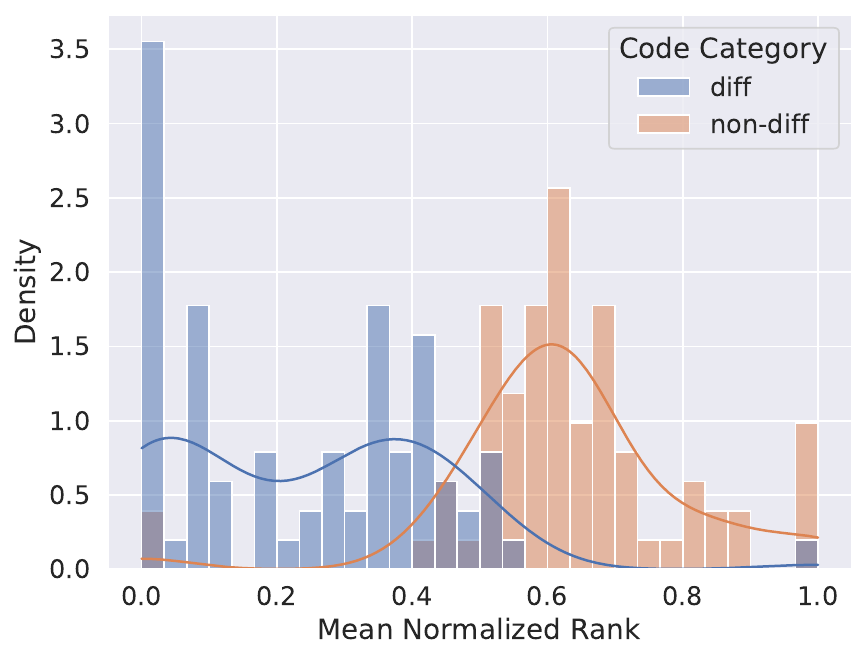}
  \caption{Deepseek on APPS}
\end{subfigure}%
\begin{subfigure}{.5\textwidth}
  \centering
  \includegraphics[width=\linewidth]{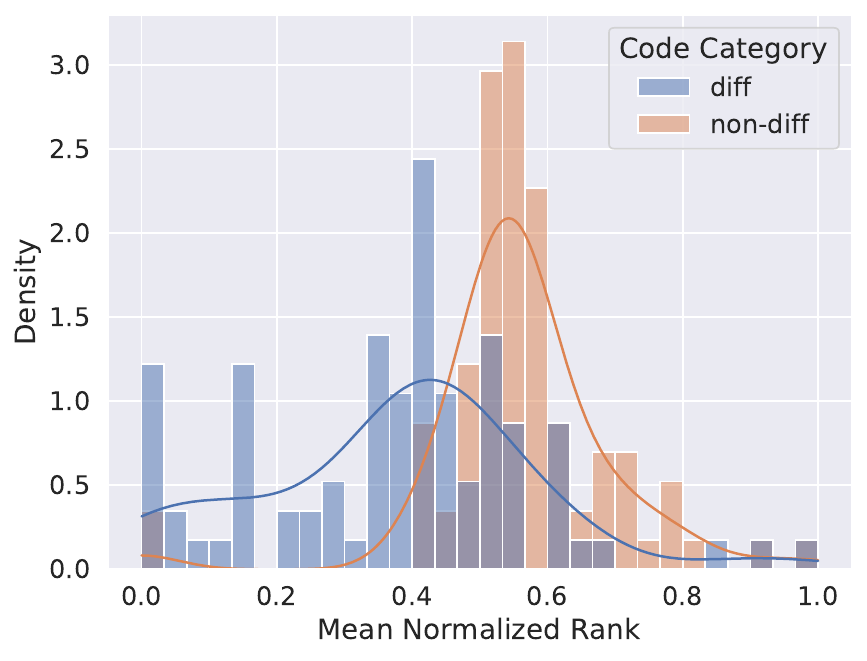}
  \caption{Llama3 on APPS}
\end{subfigure}
\end{subfigure}
\caption{Mean normalized rank distributions attribution scores that measure how much repair programs depend on self-reflections. Lower values represent higher attribution scores. Self-reflections cause targeted code diff edits.}
\label{fig:attr_sol:other}
\end{figure}
\section{More Interpretability Results}
\label{sec:interpret}
\autoref{fig:attr_feedback:other} and \autoref{fig:attr_sol:other} show mean normalized rank analyses on MBPP and APPS. We observe the same results as on HumanEval in the main paper. We further confirm that diff lines having lower mean normalized rank than non-diff lines is statistically significant by recording the p-value of one-tailed t-tests. All the p-values in \autoref{table:pv_reflection} and \autoref{table:pv_sol} are smaller than 0.05.

\begin{table}[t]
    \centering
    \begin{tabular}{cccc}
        Model & HumanEval & MBPP & APPS \\ \hline
        Deepseek & $0.001$  & $0.0016$ & $0.0024$ \\
        Llama3 & $1.59 \times 10^{-5}$ & $1.92 \times 10^{-6}$ & $7.82 \times 10^{-16}$ \\ \hline
    \end{tabular}
    \caption{P-value for one-tailed t-tests for mean normalized ranks between diff and non-diff lines. Attribution scores here measure self-reflections' dependencies on buggy programs.}
    \label{table:pv_reflection}
\end{table}

\begin{table}[t]
    \centering
    \begin{tabular}{cccc}
        Model & HumanEval & MBPP & APPS \\ \hline
        Deepseek & $2.07 \times 10^{-11}$  & $0.0002$ & $2.46 \times 10^{-25}$ \\
        Llama3 & $3.89 \times 10^{-12}$ & $5.25 \times 10^{-26}$ & $4.81 \times 10^{-11}$ \\ \hline
    \end{tabular}
    \caption{P-value for one-tailed t-tests for mean normalized ranks between diff and non-diff lines. Attribution scores here measure program repairs' dependencies on self-reflections.}
    \label{table:pv_sol}
\end{table}

\section{Prompts}
\label{sec:prompts}
\textbf{Initial greedy decoding prompt}:

Complete \{query\} Keep the function signature and docstring if there is one. Wrap your code with \texttt{\textasciigrave\textasciigrave\textasciigrave}. Do not include any assert statements.

\textbf{Self-reflection generation prompt}:

System prompt: You are experienced at identifying issues in Python programs. You are capable of summarizing issues with plain languages that are easy for language models to understand.

User: I am going to show you a Python programming exercise and a Python solution. The solution is incorrect. Your task is to describe the issues in the incorrect solution in a few sentences. Think step-by-step and pay attention to the test cases. Be as specific as you could. It is very important that you produce the best response. Start your response with \texttt{\textasciigrave\textasciigrave\textasciigrave}The solution is incorrect because\texttt{\textasciigrave\textasciigrave\textasciigrave}. Do not write any code.

The exercise is \{query\} The incorrect solution is \{program\} It failed the following test cases \{errors\}

\textbf{Program Repair prompt based on self-reflection}:

User: Complete {query}

Assistant: {program}

User: {self-reflection} Based on the feedback, correctly complete {query} Wrap your code with \texttt{\textasciigrave\textasciigrave\textasciigrave}. Do not include any assert statements.

\textbf{GPT-4 evaluate self-reflection prompt}:
I am going to show you a programming exercise and an incorrect to the exercise. Then I am going to show you an explanation about why the incorrect solution is incorrect. Your task is to classify whether the explanation correctly identifies issues in the code. Respond yes if the explanation is correct and no otherwise. Your response should only consist of one word: "yes" or "no".

The exercise is {query} The incorrect solution is {program} It failed the following test cases {errors} The explanation is {self-reflection} Please analyze the explanation carefully and respond with "yes" or "no".

\section{Compute Resources}
\label{sec:compute}
All our experiments use 4 NVIDIA A100 GPUs.

\end{document}